%% file: OSHAGH.tex
\documentclass[graybox]{svmult}

\usepackage{mathptmx}       
\usepackage{helvet}         
\usepackage{courier}        
\usepackage{type1cm}        
\usepackage{makeidx}         
\usepackage{graphicx}        
\usepackage{multicol}        
\usepackage[bottom]{footmisc}
\usepackage{natbib}

\makeindex             

\begin{document}

\input{journal_abbreviations}

\title*{Noise Sources in Photometry and Radial Velocities}
\author{Mahmoudreza Oshagh}

\institute{Mahmoudreza Oshagh \at Institut f\"{u}r Astrophysik, Georg-August-Universit\"{a}t G\"{o}ttingen, Friedrich-Hund-Platz 1, 37077 G\"{o}ttingen, Germany,\\
\email{moshagh@astro.physik.uni-goettingen.de}\\ \\
Instituto de Astrof\'isica e Ci\^encias do Espa\c{c}o, Universidade do Porto, CAUP, Rua das Estrelas, 4150-762 Porto, Portugal
}

%
%
\maketitle

\abstract{The quest for Earth-like, extrasolar planets (exoplanets), especially those located
inside the habitable zone of their host stars, requires techniques
sensitive enough to detect the faint signals produced by those
planets. The radial velocity (RV) and photometric transit
methods are the most widely used and also the most efficient methods
for detecting and characterizing exoplanets. However, presence of astrophysical ``noise'' makes it difficult to detect and accurately characterize exoplanets. It is important to note that the amplitude of such astrophysical noise is larger than both the signal of
Earth-like exoplanets and state-of-the-art instrumentation limit precision, making this a pressing
topic that needs to be addressed. In this chapter, I present a general review of the main sources of noise in photometric and RV observations, namely, stellar oscillations, granulation, and magnetic activity. Moreover, for
each noise source I discuss the techniques and observational strategies which allow us to mitigate their impact.}

\section{Introduction}
\label{sec:Intro}

Exoplanetology is a vigorous and exciting new area of astrophysics. Since the revolutionary discovery
of a planet orbiting the solar-like star 51 Peg \citep{Mayor-95}, over 3500 exoplanets have been
discovered in about 2600 planetary systems\footnote{\url{http://exoplanet.eu}}, which places our unique Solar System into context
through the new field of comparative planetology.
The radial velocity (RV) and photometric transit methods are the most widely used --- and most successful techniques --- in the detection and characterization of exoplanets. 

Exoplanet-host stars are also the source of astrophysical ``noise'' with different amplitudes and timescales that can hamper the detection of accurate numbers of planets in a system and the accurate characterization of the detected planets. In this chapter, I provide a general review of the different sources of noise which are present in high-precision photometric and RV observations. Since the timescales of these noise signals are diverse, I have thus organized this chapter in such a way that the timescales of the noise signals increase as we move along. Moreover, for each noise source I discuss several proposed observational strategies and data analysis techniques which could help eliminate their impact.

\section{Stellar oscillations}
\label{sec:Osci}
Due to the presence of pressure waves in the interiors of stars, stellar surfaces often exhibit oscillations. The typical amplitude and
timescale of oscillation modes increase with stellar mass along the main
sequence. For that reason, the measurement of stellar oscillations has been used to extract crucial information about the interior structure of stars as part of a field of stellar astrophysics known as asteroseismology. As a consequence, asteroseismology has enabled us to characterize stellar properties with extremely high precision \citep{Christensen-Dalsgaard-16}. Nevertheless, in the field of exoplanets the stellar oscillation signal is regarded as a source of noise that can hamper the detection of weak exoplanet-induced signals.

\subsection{Radial velocities}
Several studies have attempted to estimate the exact timescale and amplitude of RV signals induced by stellar oscillations. For example, \citet{Bazot-07} used the HARPS spectrograph to perform extensive high-precision RV measurements of $\alpha$ Cen A during five consecutive nights with very short exposure times. Their observations revealed the timescale of the oscillations to be of the order of 5--15 min and the RV amplitude to be in the range 0.2--$3\:{\rm m\,s^{-1}}$.

\subsubsection{Eliminating stellar oscillation noise in RV}
\citet{Santos-07} explored various
observing strategies to reduce the induced RV signal due to stellar oscillations. They reached to the conclusion that the appropriate strategy is to use long exposure times (minimum of 15 min per exposure) so as to significantly average out the stellar oscillation noise. Subsequently, this strategy has been used in performing RV measurements using stable spectrographs such as HARPS.

\subsection{Photometry}
Short-cadence and high-precision photometric observations obtained with the \textit{Kepler} space telescope \citep{Borucki-10} allowed us to estimate the timescale and amplitude of the stellar oscillation noise in photometry. The range of timescales closely matches the values obtained from RV observations (5--15 min). The amplitude of photometric variations due to stellar oscillations was seen to lie in the range 100--300 parts-per-million (ppm) \citep[e.g.,][]{Carter-12}.

\subsubsection{Eliminating stellar oscillation noise in photometry}\label{sec:elimoscphot}
Similarly to the RV case, long-exposure photometric observations would mitigate the impact of stellar oscillations. However, since transits can have durations of only a few hours, a long-exposure strategy would negatively impact on the detection of transiting exoplanets. Consequently, large surveys such as \textit{Kepler} and \textit{CoRoT} have provided high-precision photometric measurements at short cadences (60-second cadence in the case of \textit{Kepler}) in order to enable the detection of transits by Earth-size planets. Moreover, short-cadence observations provide more data points during the transit, and hence facilitate accurate estimation of a planet's parameters through the analysis of the transit light curve. Note that it is not uncommon to bin the transit light curve (with bin sizes of 15--30 minutes) in order to cancel out the effect of stellar oscillations \citep[e.g.,][]{Barclay-13}.

\section{Granulation}
\label{sec:Gran}
Stars with convective envelopes exhibit a granulation pattern at their surfaces. The granulation pattern\footnote{Granulation patterns at the surfaces of stars can only be observed through the analysis of spatially-resolved images, which are currently only possible for the Sun.} manifests itself as the upward flow of bright and hot material from deeper layers followed by a downward flow of darker
material after being cooled off at the surface. Stellar granulation adds substantial correlated noise to RV and photometric time-series observations, with larger amplitudes and longer timescales than the noise due to stellar oscillations.

\subsection{Radial velocities}
\citet{Dumusque-11} used HARPS to obtain long-term, continuous and high-precision RV measurements for five stars of different spectral types. They modeled the resulting RV power density spectra using a functional form that had been previously introduced to describe the granulation signal in the Sun \citep[the so-called Harvey-like profile;][]{Harvey-85}. As a result, \citet{Dumusque-11} estimated the timescale of the granulation noise to lie in the range from 15 minutes to 24 hours, and the amplitude to be in the range 1--$30\:{\rm m\,s^{-1}}$, depending on the spectral type of the star.

\subsubsection{Eliminating granulation noise in RV}

\citet{Dumusque-11} also evaluated several observational strategies to reduce the RV noise due to granulation. They established that the best observational strategy is to obtain three RV measurements per night for each star separated by one to two hours. They demonstrated that this approach can significantly reduce the granulation noise. This strategy has ever since been used when performing RV observations with spectrographs such as HARPS, HARPS-N, and SOPHIE.

\subsection{Photometry}
Based on solar observations obtained with the 
\textit{SOHO} spacecraft, \citet{Jenkins-02} demonstrated that the granulation of the quiet Sun can produce photometric variability of up to 50 ppm. More recently, several studies have used short-cadence photometric observations obtained with space telescopes such as \textit{Kepler} and \textit{CoRoT} and proceeded with the analysis of the corresponding power density spectra. The amplitude and timescale of the granulation noise in photometric observations across different spectral types and evolutionary
states have been constrained as a result. \citet{Gilliland-11} provide scaling relations for the estimation of the amplitude and timescale of the granulation noise in photometric observations:
\begin{equation}
\sigma_{\rm gran}=(75\:{\rm ppm}) \left( \frac{M}{{\rm M}_{\odot}}\right) ^{-0.5} \left( \frac{R}{{\rm R}_{\odot}}\right)  \left( \frac{T_{\rm eff}}{T_{{\rm eff},\odot}}\right)^{0.25}
\end{equation}
and
\begin{equation}
\tau_{\rm gran}=(220\:{\rm s}) \left( \frac{M}{{\rm M}_{\odot}}\right) ^{-1} \left( \frac{R}{{\rm R}_{\odot}}\right)^{2}  \left( \frac{T_{\rm eff}}{T_{{\rm eff},\odot}}\right)^{0.5} \, .
\end{equation}  
The granulation noise amplitude is close to the expected amplitude of a transit signal of an Earth-size planet, hence it could become a serious obstacle for the detection and
characterization of small planets via the transit method.

\subsubsection{Eliminating granulation noise in photometry}

Just as in Sect.~\ref{sec:elimoscphot}, averaging photometric observations in order to reduce the impact of granulation will directly hamper the detection of a transiting planet's signal. Therefore, light curves are ideally obtained with short cadence to ensure that no transit signals are missed. Once the transit signal has been detected, the light curve is then binned to average out the granulation noise.

\section{Stellar magnetic activity}
\label{sec:Acti}

Stellar magnetic activity manifests itself in the form of various contrasting structures at the stellar surface (e.g., dark spots and bright faculae), commonly known as stellar active regions. The combination of active regions present at the stellar surface and stellar rotation generates RV and photometric signals with amplitudes and periods commensurate with those of exoplanet-induced signals. \citet{Basri-13} found that more than 30\% of the $150{,}000$
stars observed by \textit{Kepler} possess significantly higher levels of magnetic activity than the Sun.
Therefore, one realizes how crucial it is to estimate the impact of stellar activity on exoplanet-induced signals as well as to mitigate its effect.

\subsection{Radial velocities}
Stellar active regions, due to their temperature contrast, affect the shape of spectral lines and as a consequence deform the cross-correlation function (CCF) profile. Since radial velocities are measured by fitting
a Gaussian function to the CCF, a deformation of the CCF profile may be compensated by an offset in the mean of the fitted Gaussian. Therefore, presence of active regions may lead to inaccurate and incorrect RV measurements. Due to stellar rotation, this incorrect RV estimate will exhibit a variation with a period close\footnote{Depending on the latitude of active regions and the stellar differential rotation, different active regions would induce different periodicities.} to the stellar rotation period. 

Convective motion at the stellar surface generates a net blueshifted RV signal. In active regions, however, convective motion is significantly reduced due to the presence of strong magnetic fields. The inhibition of convective blueshifts in these regions thus leads to extra RV variations \citep{Meunier-10, Dumusque-14}. 

Consequently, it is a challenging task to assure that the observed RV variations are purely due to the Doppler
reflex motion caused by the presence of exoplanets. The presence of activity-induced RV noise has been known since the very beginning of Doppler exoplanet searches \citep{Saar-97, Hatzes-99, Santos-00, Queloz-01}. Later on, with the emergence of high-precision RV measurements\footnote{A precision of $0.5\:{\rm m\,s^{-1}}$ was achieved by the HARPS spectrograph, which enabled the detection of signals due to low-mass/Earth-size planets.}, it became clearer how crucial it is to correct for the activity-induced noise in order to be able to detect the signal due to low-mass planets in the habitable zones of solar-like stars \citep{Boisse-09,Boisse-11,Dumusque-12}. Moreover, determining the exact number of planets in a system and estimating their masses has been a challenging task whenever in the presence of stellar magnetic activity. The CoRoT-7 system best demonstrates this. Depending on the methods used to model the stellar activity noise as well as on the techniques employed to disentangle the activity and exoplanet signals, several teams have obtained conflicting results on the number of planets in the system and on their masses. In Table \ref{tab:corot7}, I summarize the number of planets and their mass estimates as obtained in different studies.

\begin{table}[t]
\caption{Mass estimates of planets in the CoRoT-7 system}
\label{tab:corot7}
\begin{tabular}{p{3cm}p{3cm}p{4cm}}
\hline\noalign{\smallskip}
CoRoT-7b & CoRoT-7c & Reference \\
\noalign{\smallskip}\svhline\noalign{\smallskip}
$4.8\pm 0.8 \,M_{\bigoplus}$ & $8.4\pm 0.9 \,M_{\bigoplus}$  & \citet{Queloz-09}\\
$6.9\pm 1.4 \,M_{\bigoplus}$ & $12.4\pm 0.42 \,M_{\bigoplus}$ & \citet{Hatzes-10}\\
$7.42\pm 1.21 \,M_{\bigoplus}$ & \ldots &  \citet{Hatzes-11}\\
$2.3\pm 1.8 \,M_{\bigoplus}$ & \ldots &  \citet{Pont-11}\\
$5.7\pm 2.5 \,M_{\bigoplus}$ & $13.2\pm 4.1 \,M_{\bigoplus}$ &  \citet{Boisse-11}\\
$8.0\pm 1.2 \,M_{\bigoplus}$ & $13.6\pm 1.4 \,M_{\bigoplus}$ &  \citet{Ferraz-Mello-11}\\
$4.8\pm 2.4 \,M_{\bigoplus}$ & $11.8\pm 3.4 \,M_{\bigoplus}$ &  \citet{Tuomi-14}\\
$4.73\pm 0.95 \,M_{\bigoplus}$ & $13.56\pm 1.08 \,M_{\bigoplus}$ &  \citet{Haywood-14}\\
$5.52\pm 0.78 \,M_{\bigoplus}$ & \ldots &  \citet{Barros-14}\\
$5.53\pm0.86 \,M_{\bigoplus}$ & $12.62\pm0.77 \,M_{\bigoplus}$ & \citet{Faria-16}\\
\noalign{\smallskip}\hline\noalign{\smallskip}
\end{tabular}
\end{table}

\subsubsection{Eliminating the activity-induced signal in RV}\label{sec:elimactrv}

There are two steps in the elimination process of activity-induced signals in RV (also known as RV jitter). The first, and main, step is to assess the presence of RV jitter. The second step is to predict its signal profile and to attempt its removal from the RV measurements.

\runinhead{Activity indicators} The first type of stellar activity indicators aim at quantifying the spectral line (or the CCF) asymmetry, e.g., the full width at half maximum ($FWHM$) of the CCF \citep{Queloz-09}, the bisector span\footnote{$BIS=V_{\rm high}-V_{\rm low}$, where $V_{\rm high}$ and $V_{\rm low}$ are the velocity average of the points at the top and bottom of the CCF profile, respectively.} \citep[$BIS$;][]{Queloz-01, Santos-02}, $V_{\rm span}$\footnote{$V_{\rm span}=RV_{\rm high} - RV_{\rm low}$, where $RV_{\rm high}$ and $RV_{\rm low}$ are Gaussian fits to the upper and lower parts of the CCF, respectively.} \citep{Boisse-11}, $V_{\rm asy}$\footnote{$V_{\rm asy}$ estimates the unbalance between the red
and blue wings of the CCF.} \citep{Figueira-13}, and the bi-Gaussian method\footnote{This approach consists in fitting a Gaussian with wings characterized by two different values of the $HWHM$ (half width at half maximum) to the CCF.} \citep{Figueira-13}.

The second type of stellar activity indicators carry information directly about the magnetic activity of the star, e.g., the average magnetic field ($B$) estimated by measuring the Zeeman splitting of spectral lines \cite[e.g.,]{Reiners-12}, the Mount Wilson $S$-index\footnote{The $S$-index is based on the measurement of the emission in the cores of the Ca II H and K lines, and reflects the non-thermal chromospheric heating associated with
the magnetic field.} \citep{Wilson-78}, and the $\log (R'_{\rm HK})$ index\footnote{$\log (R'_{\rm HK})$ is closely related to the $S$-index, giving the emission in the narrow bands normalized by the bolometric brightness of the star.} \citep{Noyes-84}.

To assess whether the observed RV signal is contaminated by RV jitter, researchers usually look for a correlation between any of the above activity indicators and the RV measurements. Presence of strong correlation means that the RV measurements need to be corrected for the RV jitter, which I describe next.

\runinhead{Modeling activity} Two main approaches have been used to model RV jitter. One approach is based on using the information provided by the activity indicators and to employ empirical proxies to predict the RV jitter. This approach has been used in detecting low-mass planets around active stars, e.g., CoRoT-7 \citep{Queloz-09,Hatzes-10,Boisse-11,Pont-11,Haywood-14,Faria-16}, GJ~674 \citep{Bonfils-07}, and HD~189733 \citep{Boisse-09, Aigrain-12}.

Another approach is based on the numerical simulation of active regions at the stellar surface, including computation of all observables (e.g., activity indicators). The synthetic RVs are then simultaneously fitted to the observed RVs and to any activity indicator measurements. There are several numerical tools available to the community capable of performing this analysis, e.g., \textsc{soap} \citep{boisse-12}, \textsc{soap2.0} \citep{Dumusque-14}, and \textsc{StarSim} \citep{Herrero-16}. This approach has been used in correcting RV observations of, e.g., HD~189733 \citep{boisse-12} and $\alpha$ Cen B \citep{Dumusque-12, Dumusque-14}. Although the use of numerical simulations has been shown to be the more robust and accurate of the two approaches described, it is also the more time-consuming from a computational perspective. 

I would like to note that stellar active regions vary spatially and temporally, further evolving over several stellar rotation periods, which makes RV jitter not a strictly periodic and stable signal. Therefore, most of the correction techniques fail to explain the real observed RV jitter, which points to the necessity of developing models that take into account physical processes related to the active regions' formation and evolution. For instance, a recent effort by \citet{Dumusque-15} aimed at observing the Sun as a star with the HARPS-N spectrograph and trying to model the solar RV variation using observables that could be obtained through analysis of the resolved images of the Sun from solar satellites.

\subsection{Photometry}
The temperature contrast of active regions also produces photometric variations, which can be periodic due to the stellar rotation. This noise signal influences the detection and characterization of planets via the transit method. One can split activity-induced photometric noise into two main types depending on their source, i.e., active regions unocculted by the transiting planet during transit and occulted regions. These two types of active regions affect the transit light curve in different ways.

\subsubsection{Unocculted stellar active regions} 
Unocculted stellar active regions lead to periodic photometric modulation due to stellar rotation. The influence of such light-curve modulation on the planetary parameter estimates has been explored in several observational and simulation studies. For instance, \citet{Czesla-09} demonstrated that the planet radius can be overestimated by up to 4\%.

\subsubsection{Occulted stellar active regions} 
In case the transiting planet occults the stellar active regions, this produces anomalies in the transit light curve that
may lead to an inaccurate estimation of the planetary parameters, e.g., the planet radius and orbital inclination. Through simulations, \citet{Oshagh-13b, Oshagh-15b} showed that the planet radius can be underestimated by 5\% due to stellar active region occultation. Moreover, \citet{Oshagh-14} demonstrated that the planet radius underestimation can be as large as 10\% if the light curve is obtained at short wavelengths.

Analysis of high-precision, transit light curves allows us to accurately measure the transit times. The variation of transit times --- known as transit-timing variation or TTV --- may indicate the presence of other non-transiting planets in the system, which perturb the orbit of the transiting planet. As shown by \citet{Oshagh-13a}, however, the anomalies caused by occulted active regions can mimic a TTV signal with an amplitude of 200 seconds, similar to the TTV signal induced by an Earth-mass planet in a mean-motion resonance with a Jovian
body transiting a solar-mass star in a three-day orbit \citep{Boue-12}.

A study by \citet{Oshagh-15} also showed that the occultation by a transiting planet of a large, polar stellar spot can smear out the transit light curve. It should be noted that large, cool (dark) and long-lived stellar spots located near
the stellar rotational axis are common features in stars regardless of the
stellar rotational velocity and spectral type \citep[e.g.,][]{Strassmeier-91}. Furthermore, the occultation of active regions can affect the estimation of the spin-orbit
angle based on measurements of the Rossiter--McLaughlin effect \citep[e.g.,][and references therein]{Oshagh-16}. 

\subsubsection{Eliminating the activity-induced signal in photometry}
The most promising strategy for estimating and eliminating the impact of stellar active regions is to use state-of-the-art models --- e.g., \textsc{soap-t} \citep{Oshagh-13a}, \textsc{macula} \citep{Kipping-12}, and \textsc{spotrod} \citep{Beky-14} --- to reproduce
the noise signal generated by these regions and to subsequently remove it from the observational
data. However, this approach faces several issues. First, the models require that assumptions be made concerning the values taken by their parameters, and there exists strong degeneracy\footnote{For instance, a stellar spot's
temperature contrast and filling factor are strongly degenerate, and cannot thus be estimated
independently.} in determining the properties of the stellar active regions. Second, running numerical models is a time-consuming process from a computational perspective. Similarly to the RV jitter correction (Sect.~\ref{sec:elimactrv}), the evolution of stellar active regions makes accurate modeling of the photometric variation a challenging and difficult task. In this regard, \textsc{macula} is the only tool which takes the evolution of stellar active regions into account by implementing a linear stellar-active region evolution model.

\section{Conclusion}
\label{sec:Conclusion}

In this chapter, I reviewed the sources and characteristics of astrophysical ``noise'' signals that contaminate RV and photometric observations in exoplanet searches. These noise signals have distinct timescales and amplitudes and, therefore, the strategies and techniques used to eliminate them will differ. In Table \ref{tab:summary}, I present a summary of the timescales and amplitudes of the several noise signals described above, as well as the most efficient way of eliminating them from our observations. More details can be obtained from the slides presented at the School (available at \url{http://www.iastro.pt/research/conferences/faial2016/files/presentations/CE6.pdf}).

\begin{table}[t]
\caption{Characteristics of astrophysical ``noise'' signals in RV and photometric exoplanet searches}
\label{tab:summary}
\begin{tabular}{p{2.3cm}p{2.0cm}p{1.2cm}p{1.6cm}p{4cm}}
\hline\noalign{\smallskip}
Noise Source & Timescale & RV & Photometry & Treatment \\
 &  & (${\rm m\,s^{-1}}$) & (ppm) & \\
\noalign{\smallskip}\svhline\noalign{\smallskip}
Oscillations & 5--15 min & 0.2--3  & 100--300  & \textbf{RV:} at least 15 min exposure \

\textbf{Photometry:} binning the light curve into 15-min bins after detection of transit signal
\\

\noalign{\smallskip}\hline\noalign{\smallskip}
Granulation & 15 min to 24 hr & 1--30  & 50--500  & \textbf{RV:} three measurements per night with 1--2 hr separation and averaging them \

\textbf{Photometry:} binning the light curve into 1-hr bins after detection of transit signal\\

\noalign{\smallskip}\hline\noalign{\smallskip}
Magnetic activity & Several days  & 1--200 & 50--$10{,}000$  & \textbf{RV:} finding correlation between measured RVs and activity indicators; if any correlation found, remove RV jitter by modeling \

\textbf{Photometry:} model out-of- and in-transit portions of light curve\\

\noalign{\smallskip}\hline\noalign{\smallskip}
\end{tabular}
\end{table}

\begin{acknowledgement}
I would like to thank the members of the SOC for inviting me to the \textit{IV$^{th}$ Azores International Advanced School in Space Sciences} held in the Azores Islands, Portugal. I acknowledge funding from the Deutsche Forschungsgemeinschaft (DFG, German Research Foundation): OS 508/1-1.

\end{acknowledgement}

\bibliographystyle{apj}
\bibliography{mahlibspot}

\end{document}

%% file: journal_abbreviations.tex
\newcommand*\aap{A\&A}
\let\astap=\aap
\newcommand*\aapr{A\&A~Rev.}
\newcommand*\aaps{A\&AS}
\newcommand*\actaa{Acta Astron.}
\newcommand*\aj{AJ}
\newcommand*\ao{Appl.~Opt.}
\let\applopt\ao
\newcommand*\apj{ApJ}
\newcommand*\apjl{ApJ}
\let\apjlett\apjl
\newcommand*\apjs{ApJS}
\let\apjsupp\apjs
\newcommand*\aplett{Astrophys.~Lett.}
\newcommand*\apspr{Astrophys.~Space~Phys.~Res.}
\newcommand*\apss{Ap\&SS}
\newcommand*\araa{ARA\&A}
\newcommand*\azh{AZh}
\newcommand*\baas{BAAS}
\newcommand*\bac{Bull. astr. Inst. Czechosl.}
\newcommand*\bain{Bull.~Astron.~Inst.~Netherlands}
\newcommand*\caa{Chinese Astron. Astrophys.}
\newcommand*\cjaa{Chinese J. Astron. Astrophys.}
\newcommand*\fcp{Fund.~Cosmic~Phys.}
\newcommand*\gca{Geochim.~Cosmochim.~Acta}
\newcommand*\grl{Geophys.~Res.~Lett.}
\newcommand*\iaucirc{IAU~Circ.}
\newcommand*\icarus{Icarus}
\newcommand*\jcap{J. Cosmology Astropart. Phys.}
\newcommand*\jcp{J.~Chem.~Phys.}
\newcommand*\jgr{J.~Geophys.~Res.}
\newcommand*\jqsrt{J.~Quant.~Spec.~Radiat.~Transf.}
\newcommand*\jrasc{JRASC}
\newcommand*\memras{MmRAS}
\newcommand*\memsai{Mem.~Soc.~Astron.~Italiana}
\newcommand*\mnras{MNRAS}
\newcommand*\na{New A}
\newcommand*\nar{New A Rev.}
\newcommand*\nat{Nature}
\newcommand*\nphysa{Nucl.~Phys.~A}
\newcommand*\pasa{PASA}
\newcommand*\pasj{PASJ}
\newcommand*\pasp{PASP}
\newcommand*\physrep{Phys.~Rep.}
\newcommand*\physscr{Phys.~Scr}
\newcommand*\planss{Planet.~Space~Sci.}
\newcommand*\pra{Phys.~Rev.~A}
\newcommand*\prb{Phys.~Rev.~B}
\newcommand*\prc{Phys.~Rev.~C}
\newcommand*\prd{Phys.~Rev.~D}
\newcommand*\pre{Phys.~Rev.~E}
\newcommand*\prl{Phys.~Rev.~Lett.}
\newcommand*\procspie{Proc.~SPIE}
\newcommand*\qjras{QJRAS}
\newcommand*\rmxaa{Rev. Mexicana Astron. Astrofis.}
\newcommand*\skytel{S\&T}
\newcommand*\solphys{Sol.~Phys.}
\newcommand*\sovast{Soviet~Ast.}
\newcommand*\ssr{Space~Sci.~Rev.}
\newcommand*\zap{ZAp}